# Spectral modification of mode structures in silver nanoparticle doped Rhodamine 6G

Arindam Sarkar, C.L.Linslal, V. P. N. Nampoori

*International School of Photonics, Cochin University of Science and Technology. Cochin, Kerala, India*
E-mail: arindam_sar@yahoo.com

*Abstract*— we show in our work that very narrow(FSR~0.09 nm) lasing modes can be formed from Rhodamine 6G solution confined within quartz($SiO_2$) cuvette with suitable pumping scheme by Q switched Nd:YAG laser. With introduction of silver nanoparticles of different concentrations in Rhodamine 6G we show that such lasing modes can be modulated as well as tuned in intensity, band spacing and emission wavelength. We also show that this system maintains a very high Q value $> 6.4*10^3$ irrespective to change in other parameters.

I. INTRODUCTION

In general dye lasers are not new. Even back in 1970s there are different examples of lasers using dyes [1-3]. Since then different schemes of lasing got developed using dyes including dye doped fiber lasers. In later days one of the approaches was developed to introduce film containing dye molecules confined in a Fabry-Perot cavity [4]. One of the major requirements is still narrow line width and easy tunability. Use of tunable dye lasers we can find even nearly forty years back [5]. Different sophisticated techniques are developed for laser tuning including micro-machined tunable coupled-cavity laser [6] and two-stage self-coupled optical waveguide (SCOW) [7]. However, there are plenty of scope still remains in the subject achieving these requirements using some simpler and low cost technique. It is also known that specific metallic geometries when close to fluorophore can modify spectral properties and can increase quantum yield [8]. In the presence of metal nanoparticles complete quenching and extensive enhancement is also demonstrated [9]. Fluorescence enhancement and nonlinear property modifications of dyes with silver nanoparticles (AgNPs) are also demonstrated very recently [10, 11]. It gives new ways of spectral tuning and modification of dye lasing in presence of metal nanoparticles near fluorophore.

## II. EXPERIMENTS

We have prepared silver nanoparticles in polyol method [12], which is very simple, low cost and reliable method producing silver nanoparticles (AgNP) up to a certain dimension of the particle. In this process ethylene glycol (EG) (99% pure) is allowed to boil at 120 degree centigrade. This temperature is maintained for five minutes when silver nitrate (pure) and polyvinyl pyrrolidone (PVP) (Mw=10000) (99 %) (Sigma-Aldrich) are added. PVP acts as a capping agent and prevent agglomeration in AgNPs. We maintained 12:1 molar ratio between $AgNO_3$ and PVP. Particles were deposited using centrifuge and collected after washing with ethanol. AgNPs were characterized using high resolution transmission electron microscopy (HRTEM) and X-ray diffraction (XRD). TEM was taken on JEOL 3010. XRD was taken on AXS Bruker D% diffractometer using Cu Kα-radiation ($\lambda=0.1541$ nm).

Rhodamine 6G ($C_{28}H_{30}N_2O_3HCl$) is a highly photo stable dye with very high quantum yield (nearly 0.95) [13]. Since it has good absorption nearly at wavelength 532 nm it is often pumped by frequency doubled Nd:YAG to produce lasers . The dye is very low cost in nature. Lasing capability of this dye was known even twenty years back from present day [14]. At low concentration the quantum yield is almost independent of concentration. Such behaviour is exhibited up to a concentration of $2*10^{-4}$ M concentration [15]. In our experiments we have maintained dye concentration of $5*10^{-5}$ M/L. We have taken absorption spectra of both AgNP and dye using spectrometer (Jasco V-570 UV/VIS/IR).

In our experiments we have taken rhodamine solution and rhodamine with different concentrations of silver nanoparticles (AgNPs) in a quartz ($SiO_2$) cuvette. This cuvette had walls of nearly 1.125 mm thick and spacing between two opposite walls was 1.025 cm. Pumping and emission collection direction are shown in Fig-1.In this present discussion we will show different results with the help of four samples i.e. pure rhodamine sample S1 and AgNP doped rhodamine colloid samples i.e S2, S3, S4 with AgNP concentration $8.4*10^{-6}$ M/L, $8.4*10^{-5}$ M/L and $6*10^{-4}$ M/L respectively. In all the cases (i.e. S1, S2, S3 and S4) ethylene glycol (EG) was taken as solvent. We have used transverse pumping to the dye using two different schemes with Q switched Nd:YAG (Spectra Physics LAB-1760, 532 nm, 7ns, 10 Hz). In the scheme-1 we have used the beam diameter of 0.6 cm to pump the system. In the scheme-2 we introduce a cylindrical lens to focus the beam horizontally in the form of a strip on the cuvette wall-1 surface. The length of the strip was also measured as nearly 0.69 cm. This scheme looks slightly different from the first but it has some interesting implications. First of all since it concentrates energy in a much narrow width it has very high fluence (optical energy per unit area) on the sample. Secondly, it creates an approximate uniform light plane through the samples we have studied. The emission was recorded from wall-2 by using a collector fiber coupled to a CCD based monochromator/spectrograph (Spectra Pro 500i) with a resolution of 0.03 nm. Collection fiber was kept with an angle approximately $30°$ with the plane of illumination.

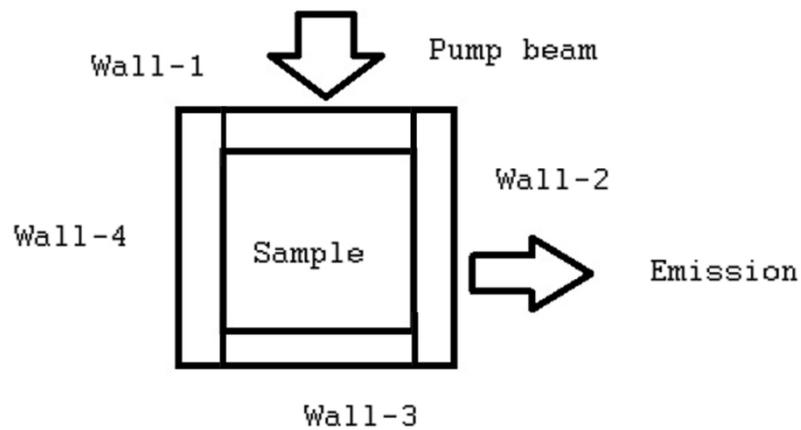

**Fig-1** Pumping and emission directions are shown schematically in the picture (top view) when sample is confined within the quartz ($SiO_2$) walls

## III. RESULTS AND DISCUSSIONS

Fig-2(a, b) shows TEM images of silver nanoparticles. Size calculated from TEM was 39.6 nm. Size distribution in the sample can be understood by Fig-2(c). Fig-2(d) is showing XRD results of a thin film made from AgNPs. In XRD we got peak at 2θ=39.147°. We can easily use Bragg's law [16] i.e. 2dSinθ=nλ and can calculate spacing between planes (d). In this relation θ is the scattering angle, λ be the wavelength of the X-ray, and n is any integer. With this d becomes 2.298 A°. We take our lattice constant a=4.0867 A° assuming face centred cube (FCC) [17]. Using $(1/d^2)=(h^2+k^2+l^2)/a^2$ with Miller indices (h,k,l), we understand that our peak belongs to the plane (1,1,1). This also matches to the JCPDS data (file No. 04-0783). Absorption spectra of AgNP sample is shown in Fig-3(a) which has a peak at 420 nm. Fig 3(b) shows absorption spectra of Rhodamine 6G (Rh6G) which we got a peak at 534 nm.

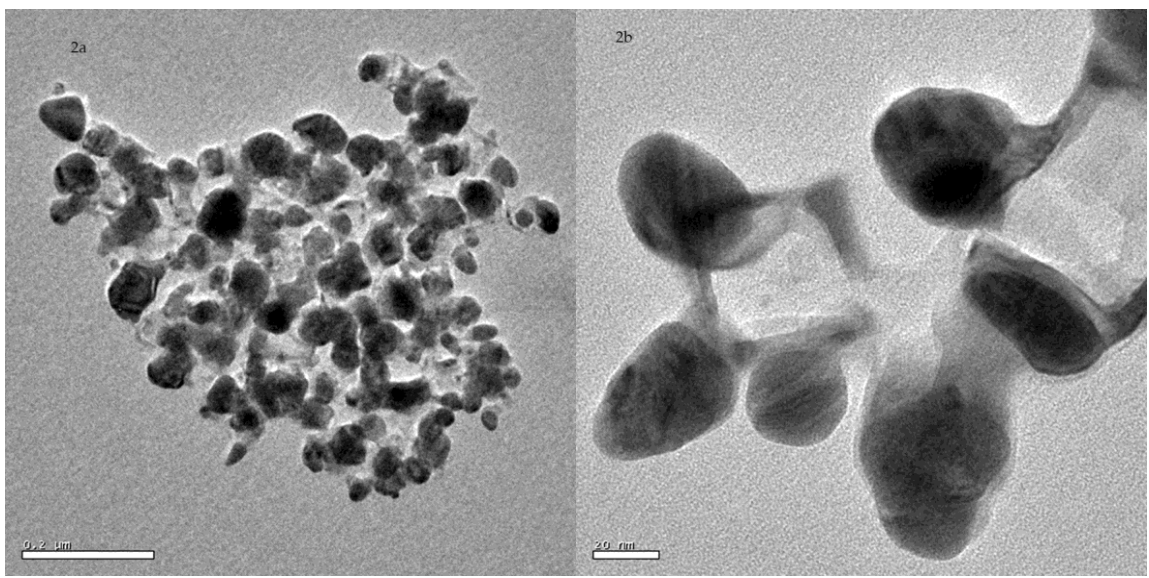

**Fig 2(a) (b)** TEM images of silver nanoparticles

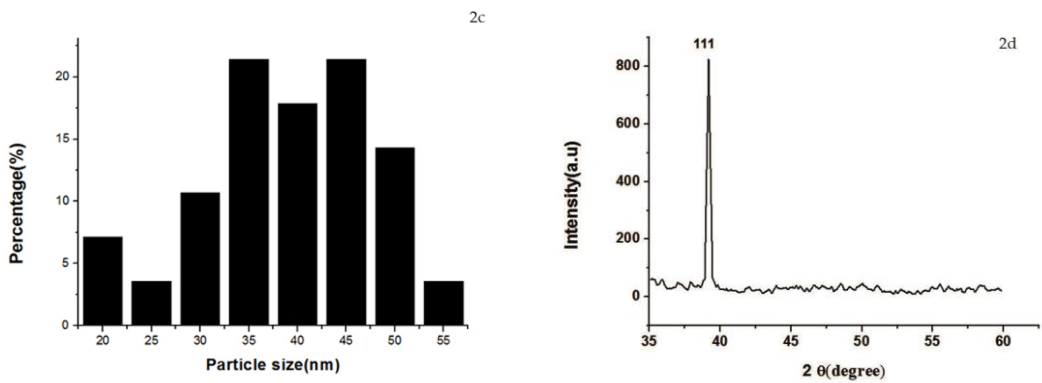

**Fig 2(c)** Size distribution of silver nanoparticles **2(d)** XRD of AgNP sample

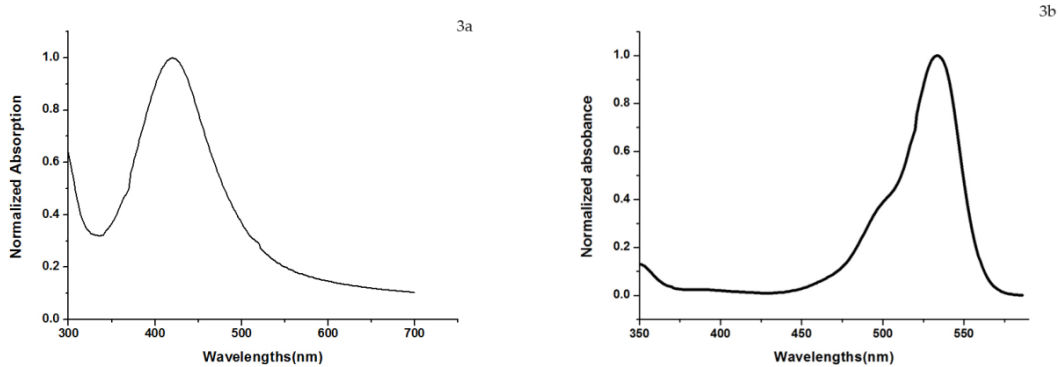

**Fig 3** Absorption spectra of **(a)** AgNP **(b)** Rhodamine 6G

Fig-4(a) shows the fluorescence of sample S1 with different energy of excitation with scheme-1. With higher energy excitation the peak fluorescence intensity of Rh6G shifts higher. Now, we have shown effect of silver concentrations in Fig-4(b). Up to concentration if we increase AgNP concentration the fluorescence gets higher for Rh6G. It even reaches to a level (as in case of S3) which 14.72 times higher than that of Rh6G alone. Metal particles generally do have very high scattering and while interacting with light scattering cross section can be much higher than that of physical cross section.

For spherical metal particles absorption cross section ($C_{abs}$) and scattering cross section ($C_{sca}$) are given by for wavelength (λ) [18]:

$$C_{abs} = \left(\frac{2\pi}{\lambda}\right) Im[\alpha], C_{sca} = \left(\frac{1}{6\pi}\right)\left(\frac{2\pi}{\lambda}\right)^4 |\alpha|^2 \tag{1}$$

α is polarizability of the particle and if we consider spherical shape sub-wavelength size particle then it is given by,

$$\alpha = 3V(\varepsilon - 1)/(\varepsilon + 2) \tag{2}$$

Where V is the volume and ε be the dielectric constant of the metal.

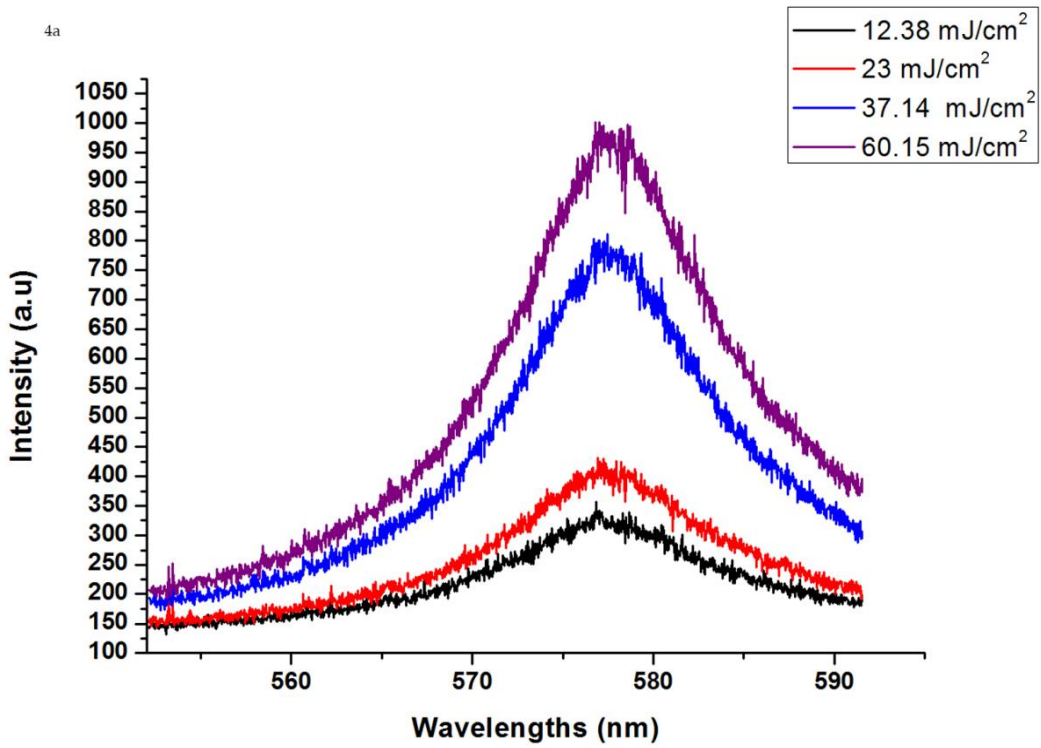

4a

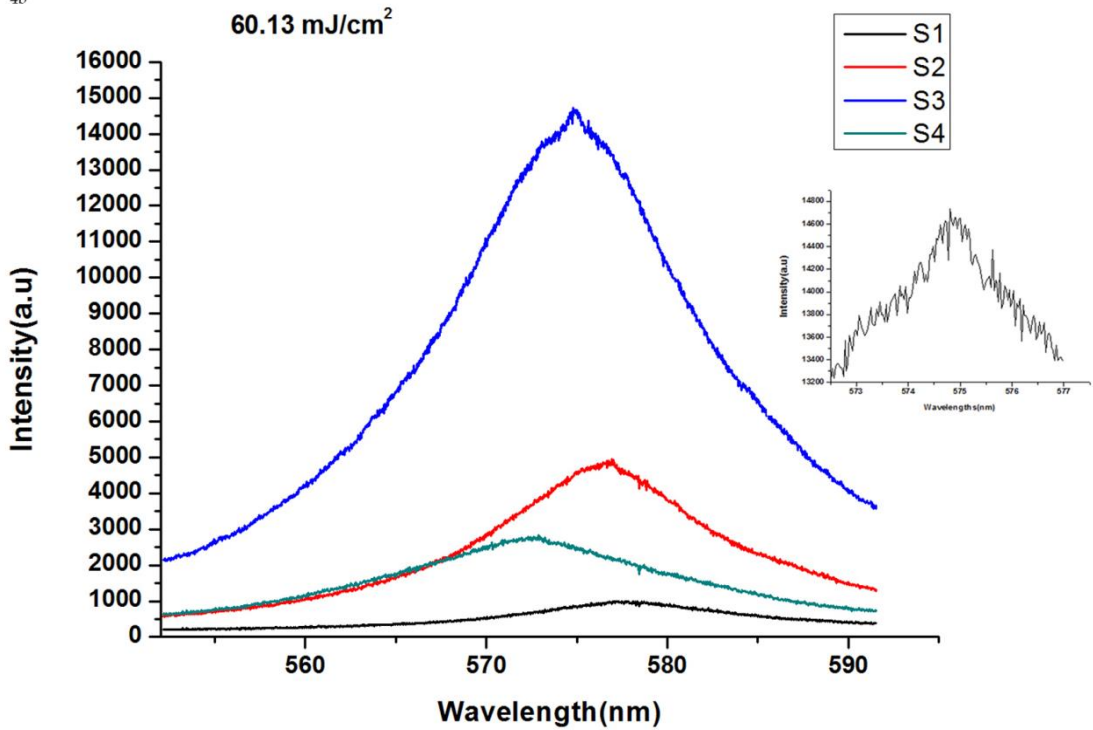

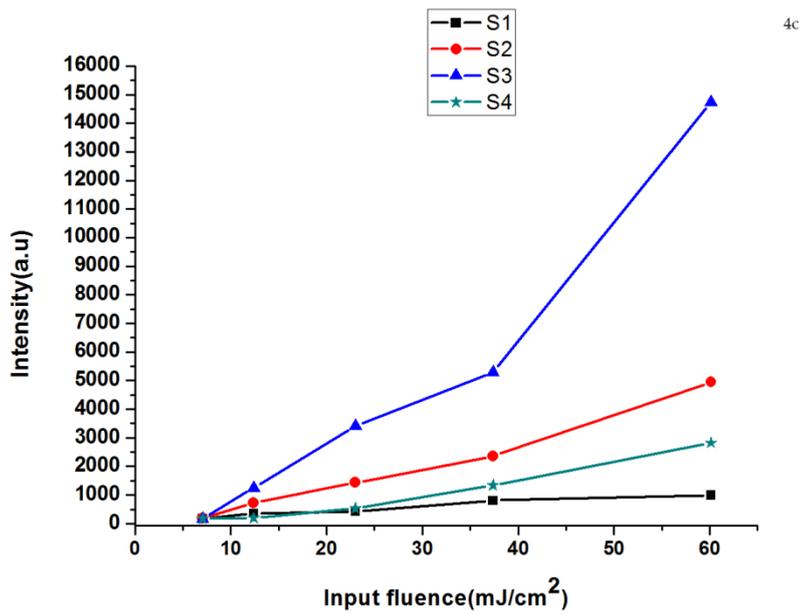

**Fig 4** showing fluorescence of **(a)** S1 for different fluence **(b)** S1, S2, S3, S4 at 60.13 mJ/cm$^2$, inset picture shows random peak appearing for S3 at 60.13 mJ/cm$^2$ **(c)** Fluorescence intensity peak variation for different samples with different input fluence

Now another consequence we find that the peak undergoes a blue shift with increase of silver concentrations. If we observe equation (1) we find that scattering cross section varies with ($1/\lambda^4$) but absorption cross section varies with ($1/\lambda$). So, in shorter wavelength it is expected to scatter more light by metal particles than that in longer wavelengths. Full width half maximum (FWHM) for Rhodamine 6G with excitation 60.15 mJ/cm$^2$ is nearly 18 nm (from 569-587 nm) with a peak at 576.8 nm. So, with increase of the silver nanoparticles it is expected that the peak will shift towards 569 nm. With $6*10^{-4}$ M/L of AgNP this peak was shifted to 572.9 nm. Though due to heavy concentration of silver after certain point quenching started to take place since total absorption of the AgNPs increase considerably and quenching becomes quite effective. But even in that case with very high concentration as in S4 the output fluorescence was more than the Rh6G alone. Now for sample S3 in high power other than enhancement we observe certain random peaks which we have shown in the inset figure of 4(b). These random peaks can be formed by the various reasons including random lasing or initial modal formation of the quartz wall of cuvette which will be discussed in detail in scheme-2. These peaks separation varied from .088 nm to 0.3 nm. Now, with Fig-4(c) we can observe how fluorescence peak variation for different samples with different input fluence. We can clearly see how after certain threshold the slope changes very large angle for S3. In a disordered media light gets scattered and travels randomly. With higher concentration of particles and higher fluence it is possible that the gain length becoming nearly equal to the length of the light path. In that case there is high probability of generation of second photon by the first photon by stimulated emission before leaving the media completely. In such scattering system several small cavities can be formed and they can act as a closed loop. If in such closed loop gain overcomes losses for a particular level of light amplification this system can act as laser resonator. It can also be noted that other than above effect emission rate from fluorophores can itself be influenced by presence metal nanoparticles near it [8]. In case of S3 since quenching is not as high as S4 and fluorescence starts getting amplified considerably at higher fluence. Now, we also showed already random peaks appearing in Fig 4(b), though it can be noted that since those peaks do not change drastically in amplitude for increase of fluences we operated we can at least ignore random lasing with resonant feedback for these input energy levels in scheme-1[19].

Now, we will move on to our discussions based on scheme-2.In this architecture dye plane in cuvette is under a very high fluence due to our excitation with narrow horizontal strip with small length. In this case the fluorescence produced by the dye gets reflected back by the cuvette wall surface partially. Since fluence was high this partial reflected optical energy is also considerably high. So, dye medium between two parallel walls can act as a gain medium at higher fluences. We observe that after a certain threshold new lasing peaks gets visible even without presence of silver or any nanoparticles. Same is shown in Fig 5(a). These peaks are very narrow and if we observe more precisely into it, we find different narrow modal structures present in each such peaks. The average intensity of modes starts increasing with increase of pumping power. Same is shown in Fig 5(b). Cases discussed here though shows only two bands but it is possible to get more number of bands with increase of pumping power to very high level. In this present article we will restrict our discussions with respect to two bands only.

Since, these periodic modal structures arise even without any nanoparticles we will first analyse the origin of such structure. Observing the nature of these modal peaks we can understand that these peaks are surely formed in an optical resonator which is part of cuvette. These peaks observed were very narrow and nearly 0.09 nm. In this case narrow modal peaks may appear due to presence of each quartz wall plate of small but finite thickness. We verified this possibility by assuming this quartz plate (e.g. plate in wall 2) is acting as a Fabry-Perot etalon.

For Fabry-Perot etalon the free spectral range (FSR) is given by [20],

$$\Delta\lambda = (\frac{\lambda^2}{2nl}) \qquad (3)$$

Where λ is wavelength, n is refractive index of of the media and l is plate thickness. Peak λ we found in our earlier experiment stated in scheme-1 was nearly 576.8 nm at 60.15 mJ/cm$^2$ for S1. The refractive index of quartz at that wavelength is nearly 1.5448 [21]. Now, by this if we calculate FSR we get a value of 0.0957 nm which is very close to the experimental value.

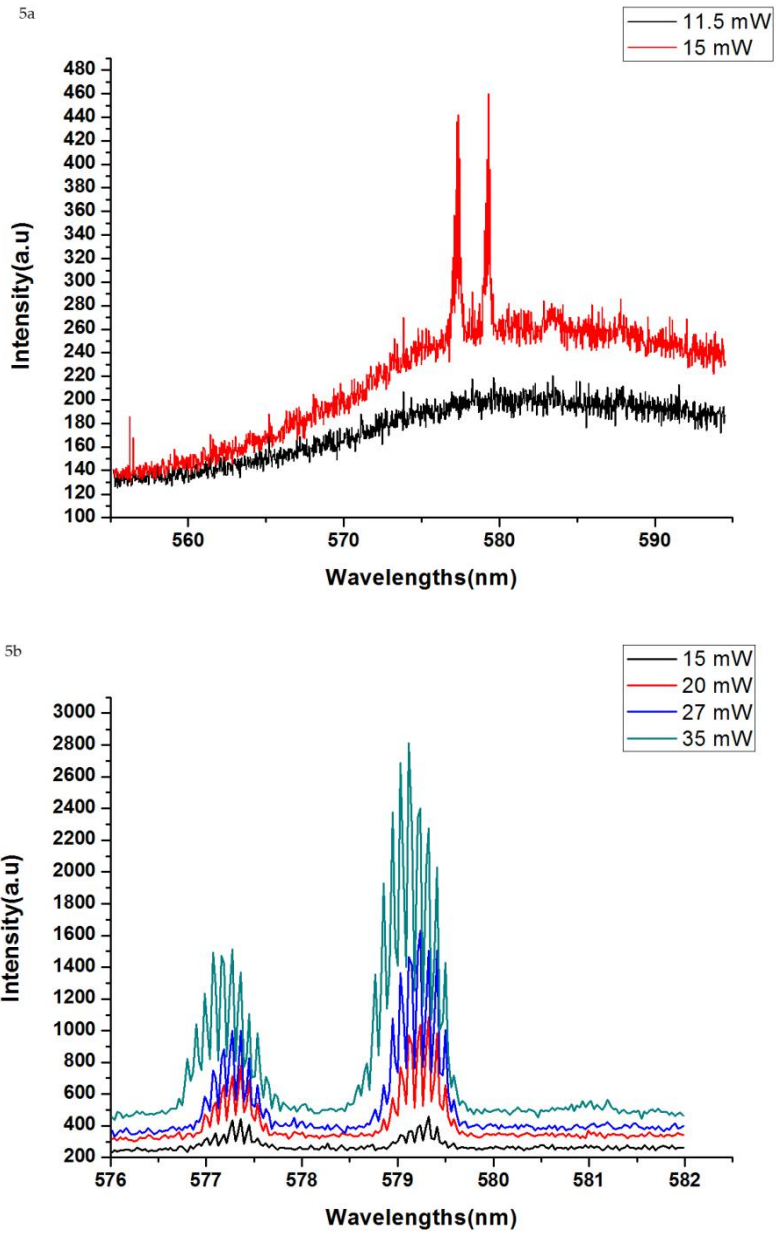

**Fig 5(a)** shows start of lasing for sample S1 after and before threshold pumping power **5(b)** precise view of narrow modal structure within lasing band for sample S1 for four different pumping powers

Now, we also observe the separation of bands nearly 1.95 nm. This can be explained by the similar reason explained by Frolov et al while discussing conducting polymer coated optical fiber [22]. In our case also two different sets of resonant frequency may appear. Refractive index of the part of the excited plane of the dye is different from the index of rest of the plane since dye index can change with high fluence by the laser. The amplified signal formed in the gain medium travels towards the wall of cuvette and we can get a new set of resonant frequencies in the dye layer near the surface of quartz plate. When optical energy enters into the wall due to the Fabry-Perot operation we explained earlier we get new sets of resonant frequencies forming whispering gallery modes (WGM). If there is a match between these two sets of resonance frequencies we will get quality factor (Q) [23] value increase and if they do not match Q factor decrease [22]. So, this creates a similar period of modulation in the system described by us.

We can write period of modulation as [22],

$$\delta\lambda = (\Delta\lambda)\left(\frac{n_{eff}}{\Delta n_{eff}}\right) \qquad (4)$$

Where $\Delta\lambda$ is actually our FSR, $\Delta n_{eff}$ is refractive index difference and $n_{eff}$ is refractive index of quartz ($SiO_2$) plate. We calculated $\Delta n_{eff} = 0.078$ for our sample S1, theoretical FSR as 0.0957 nm, n=1.5448 then $\delta\lambda$=1.895 nm which is very close to our experimental value.

Up to this point whatever we have discussed for scheme-2 was about S1 which is Rh6G sample in EG. Now, we will describe the changes undergo when AgNPs are mixed in the dye with different concentrations. Fig-6 shows how spectrum of rhodamine changes when higher concentrations of AgNPs are present. In this figure we observe three primary changes in the spectrum. First of all, up to certain concentration the average peaks starts increasing and if we go further beyond that point it starts decreasing again for a particular pumping power. It has similar reason as we mentioned while explaining enhancement and quenching effects undergone during scheme-1. Fig-7(a) shows the average modal intensity for the samples at different pumping powers. We can observe that up to sample S3 the intensity change per unit change in power is increasing with increase in concentrations. For S4 the average modal intensity is decreased but still it is higher than that of low concentration AgNP sample S2. In case of S3 this change is highest and increases almost in a perfect linear manner. For higher fluence the slope of change in average intensity for S4 is almost parallel to S3. Fig 7(b) shows maximum modal peak intensity for the samples at different pumping powers. If we consider S3 we can note that after enhancement to a certain level the maximum peak intensity value, beyond that peak intensity starts saturating. But if we consider S4 where much higher AgNPs present we can find that after a certain threshold the maximum peak intensity as well as average peak intensity starts rising much sharply compared to other samples. This is probably due to crossing the barrier of quenching effect at higher power by considerable light amplification in the cavity. As second observation we find that in Fig-6, samples also undergoes similar blue shift as we observed in scheme-1 with higher concentrations. If we consider blue shift between sample S1 and S4 it is nearly 5.7 nm.

As per third observation we also find that the band spacing between the samples decreases with increase in AgNP concentrations. For S2 it is nearly 1.93 nm and for S3 it is nearly 1.85 nm. This is mainly controlled by two different effects both dependent on AgNP concentrations. We know silver has very low refractive index near 575 nm which is as low as 0.12 [24]. So, presence of AgNPs can decrease effective refractive index of this colloidal solution and which means $\Delta n_{eff}$ starts increasing which causes $\delta\lambda$ value to fall. Increase of silver nanoparticles in the solution also can enhance the weak resonant frequencies which also can broaden each band and hence can reduce spacing between them. In our experiments we found that if we continue to increase the concentration of AgNPs then this band pattern is disrupted forming a single band with much higher width. It can be easily seen in sample S4 how modal structure has changed forming a single band in Fig-6. If we consider width of this band at 35 mW it is nearly 1.6 nm compared to S1 with nearly 0.73 nm and S3 with 0.79 nm. But even though bands got almost merged in case of S4 if we carefully look into the modal pattern formed we can clearly observe signature of more than one band. Inset picture in Fig-6 shows the line joining the lowest point of either side of the bands and modal peak points. In this figure it can be seen that that three different parts appearing in a single band. The spacing between right and left part peaks is nearly 0.91 nm. Other than the two reasons for band spacing tuning discussed by us there can be one more effect involved in it. At higher concentrations of AgNP it is possible to form large number of small resonator cavities which can produce new lasing modes due to random lasing. Though in lower input power it is likely that very low energy enters into cuvette wall due to quenching and which can result very low number of modes. For example, at 15 mW pumping we could find only one mode formed for S4. When at higher pumping light amplification in the medium is more than the quenching effect such peaks starts getting visible. So, due to these three effects the resonance frequencies in the colloid media can be influenced by AgNP with different concentrations. This perhaps explains the changing band spacing pattern for doped samples with different AgNP concentrations.

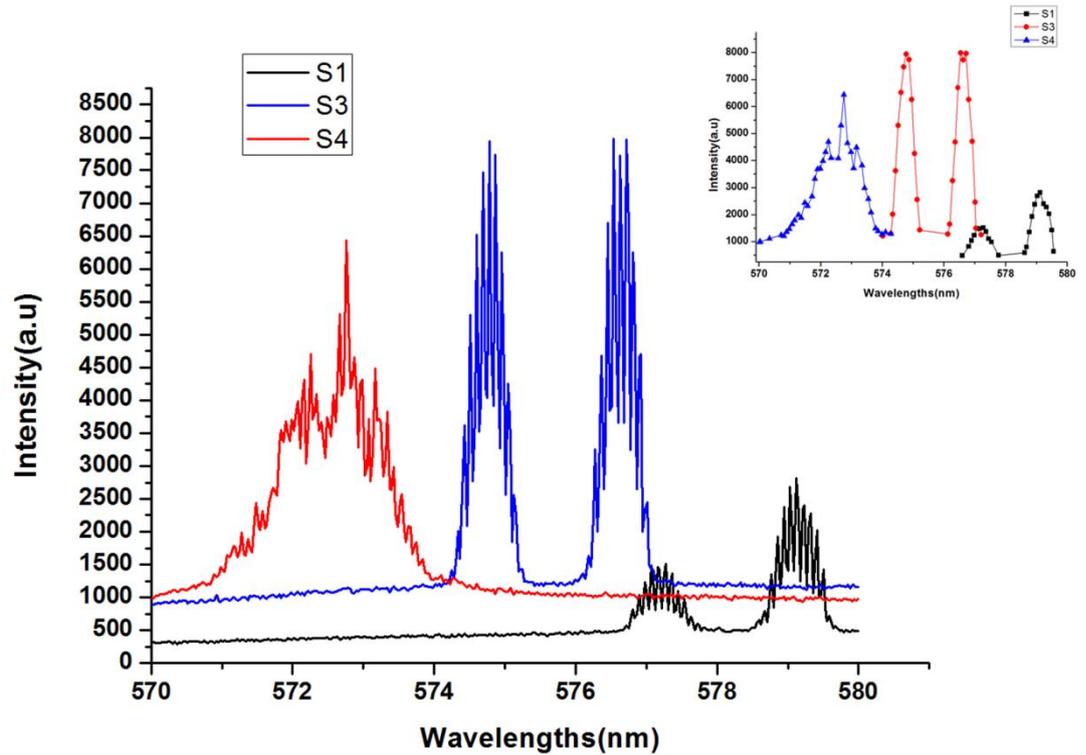

**Fig 6** shows how introduction of AgNPs change the spectrum and modal structure for different samples at 35mW. Inset picture shows the lines joining lowest point of either side of the band and all modal peaks for different samples at 35 mW.

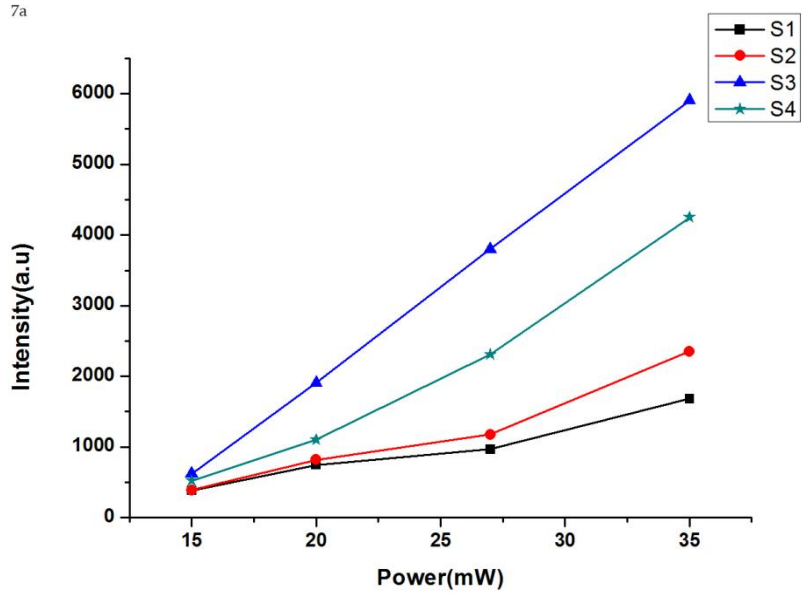

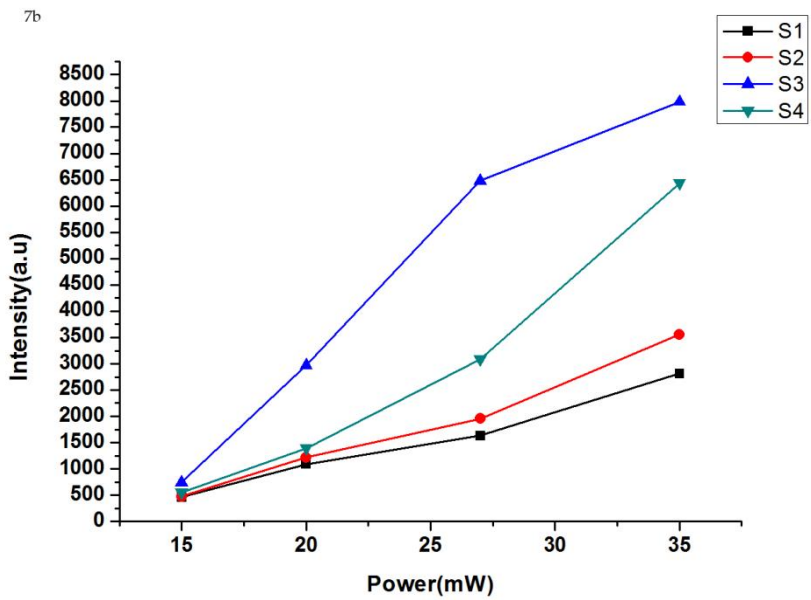

**Fig 7 (a)** Average modal intensity for different pumping powers for S1, S2, S3, S4 **(b)** Maximum value of modal intensity peak for S1, S2, S3, S4 for different pumping powers

Still now we have discussed about modification and tailoring of spectrum with respect to frequency shift, band period adjustment and intensity of the modes. But in all cases the average FSR remains almost same since it is governed by equation (3) and primarily dependent on wavelength, dimension of quartz plate and refractive index of the same. For all the samples it remains nearly 0.09 nm with variation of +/- 0.002 nm for different power. This means a stable Q value over $6.4*10^3$ is maintained irrespective of silver concentrations, band spacing and pumping power. It is notable that we achieved an extensive enhancement where the highest enhancement orders of a particular mode (~2.9 times at 35 mW comparing S1 and S3) is higher compared to the similar recent work [10].

In recent past there are different works with metal nanoparticles for enhancing the intensity of light [9-11]. But it remains difficult to achieve a stable high Q emission for individual modes over a broad range. In our simple approach we are not only achieving modal enhancement of individual peaks to a desired level but also keeping a stable Q value over a broad range of modes. In our work we also show some interesting new perspective of silver nanoparticle based tuning such as shifting of bands and tuning of band spacing. Since this approach can produce bands separated by ~ 1.8-1.9 nm such system can easily be coupled with an optical transmitter receiver either directly or via a modulator section.

## IV. CONCLUSION

In this present article we showed how we can generate very high intense and narrow lasing modes with introduction of silver nanoparticles in Rhodamine 6G. We can tune it to a desired modal intensity as this intensity amplitude varies with AgNP concentration and this enables an easy modal enhancement technique. These modes can be easily collected by a fiber from either side of the cuvette. We also show how period of modulation and spectrum shift is possible with introduction of AgNP with different concentrations. These properties can be very useful for tuning, modulation and desired spectrum tailoring of lasing modes. We have demonstrated this in a simple optical resonator however this concept can be extended further for different optical resonators. For example dye containing AgNP can be doped into polymer which can be coated to an optical fiber and can be transversely pumped for lasing. Hence, the demonstration and concept used by us can be useful for different relevant and similar applications. This tuning and modification method is comparatively low cost and maintains a very high Q value which perhaps is the most important prospect with context of our work.

ACKNOWLEDGEMENT

Authors wish to acknowledge SAIF-KOCHI for XRD facility and SAIF, IIT Madras for TEM facility. Author AS likes to thank Mathew S for his constant encouragement and support.